\documentclass[11pt,a4paper]{article}
\usepackage[english]{babel}
\usepackage{amsmath,amsthm,amssymb,epsfig,latexsym}
\usepackage{color}
\usepackage{ulem}
\usepackage[numbers,square,sort&compress]{natbib}

\newcommand{\so}{\scriptscriptstyle \rm I}
\newcommand{\st}{\scriptscriptstyle \rm I\hspace{-1pt}I}

\newcommand{\bt}{\bar t}
\newcommand{\bx}{\bar x}
\newcommand{\bs}{\bar s}
\newcommand{\by}{\bar y}
\newcommand{\bz}{\bar z}
\newcommand{\bom}{\bar\omega}
\newcommand{\uc}{u^{\scriptscriptstyle C}}
\newcommand{\ub}{u^{\scriptscriptstyle B}}
\newcommand{\vc}{v^{\scriptscriptstyle C}}
\newcommand{\vb}{v^{\scriptscriptstyle B}}

\newcommand{\bu}{\bar u}
\newcommand{\bv}{\bar v}

\newcommand{\buc}{\bar{u}^{\scriptscriptstyle C}}
\newcommand{\bub}{\bar{u}^{\scriptscriptstyle B}}
\newcommand{\bvc}{\bar{v}^{\scriptscriptstyle C}}
\newcommand{\bvb}{\bar{v}^{\scriptscriptstyle B}}
\newcommand{\bucb}{\bar{u}^{\scriptscriptstyle C,B}}
\newcommand{\bvcb}{\bar{v}^{\scriptscriptstyle C,B}}

\newcommand{\bet}{\bar\eta}

\newcommand{\bw}{\bar w}

\newcommand{\be}[1]{\begin{equation}\label{#1}}
\newcommand{\ba}[1]{\begin{multline}\label{#1}}
\newcommand{\ee}{\end{equation}}
\newcommand{\ea}{\end{eqnarray}}

\newcommand{\num}{\\\rule{0pt}{20pt}}

\newcommand{\diag}{\mathop{\rm diag}}

\newcommand{\str}{\mathop{\rm str}}

\newtheorem{prop}{Proposition}[section]
\newtheorem{lemma}{Lemma}[section]

 \makeatletter
 \@addtoreset{equation}{section}
 \makeatother

\newcommand{\bea}{\begin{eqnarray}}
\newcommand{\eea}{\end{eqnarray}}

\newcommand{\mb}[1]{\quad\mbox{#1}\quad}

\begin{document}

\begin{flushright}
LAPTH-028/16
\end{flushright}

\vspace{22pt}

\begin{center}
\begin{LARGE}
{\bf   Scalar products of Bethe vectors \\[2mm] in models with $\mathfrak{gl}(2|1)$ symmetry\\[2mm]
2. Determinant representation}
\end{LARGE}

\vspace{40pt}

\begin{large}
{A.~Hutsalyuk${}^{a}$,  A.~Liashyk${}^{b,c}$, S.~Z.~Pakuliak${}^{a,d}$,\\[1ex]
 E.~Ragoucy${}^e$, N.~A.~Slavnov${}^f$  \footnote{%
hutsalyuk@gmail.com, a.liashyk@gmail.com, stanislav.pakuliak@jinr.ru, eric.ragoucy@lapth.cnrs.fr,\\ nslavnov@mi.ras.ru}}
\end{large}

 \vspace{12mm}

${}^a$ {\it Moscow Institute of Physics and Technology,  Dolgoprudny, Moscow reg., Russia}

\vspace{4mm}

${}^b$ {\it Bogoliubov Institute for Theoretical Physics, NAS of Ukraine,  Kiev, Ukraine}

\vspace{4mm}

${}^c$ {\it National Research University Higher School of Economics,  Russia}

\vspace{4mm}

${}^d$ {\it Laboratory of Theoretical Physics, JINR,  Dubna, Moscow reg., Russia}

\vspace{4mm}

${}^e$ {\it Laboratoire de Physique Th\'eorique LAPTH, CNRS and Universit\'e de Savoie,\\
BP 110, 74941 Annecy-le-Vieux Cedex, France}

\vspace{4mm}

${}^f$ {\it Steklov Mathematical Institute of Russian Academy of Sciences, Moscow, Russia}

\end{center}

\vspace{1cm}

\centerline{ Dedicated to the memory of P.P. Kulish}

\vspace{4mm}


\begin{abstract}
We study  integrable models with $\mathfrak{gl}(2|1)$ symmetry and solvable by nested algebraic Bethe ansatz.
We obtain a determinant representation for scalar products of
Bethe vectors,  when the Bethe parameters obey some relations weaker than the Bethe equations. This representation allows us
to find the norms of on-shell Bethe vectors and obtain determinant formulas for form factors
of the diagonal entries of the monodromy matrix.
\end{abstract}

\vspace{4mm}


\vspace{1cm}

\newpage

\section{Introduction}

The present paper is a continuation of the work \cite{HutLPRS16c}, where  we began to study the problem of calculating scalar products of Bethe vectors in
$\mathfrak{gl}(2|1)$ integrable models solvable by the algebraic Bethe Ansatz. There we considered the scalar product of generic Bethe vectors and obtained
a sum formula for it.
By generic Bethe vectors we  mean
that the Bethe parameters are supposed to be generic complex numbers. In this paper we consider the case where the Bethe parameters satisfy certain
conditions, closely related to the Bethe equations but less constrained than them.  This allows us to derive a determinant representation for the scalar product.

The problem of computing the scalar products is of great importance in the algebraic Bethe ansatz \cite{FadST79,BogIK93L,FadLH96}. This is a necessary tool for
calculating form factors and correlation functions within the framework of this method. The first results in this field concerning $\mathfrak{gl}(2)$-based models and their $q$-deformations were  obtained in \cite{Kor82,IzeK84,Ize87}. This explicit representation (Izergin--Korepin formula) was, however, rather
inconvenient for applications. The next step was done in \cite{Sla89}, where a determinant representation for a particular case of scalar products
was found. In this particular case one of the vector remains generic, while the other vector becomes on-shell (that is, its Bethe parameters satisfy the
Bethe equations). The determinant representation for the scalar products opened a way for studying form factors and correlation functions in the
models with $\mathfrak{gl}(2)$ symmetry  and their $q$-deformations \cite{KitMT00,KitMST02,KitKMST09b,GohKS04,GohKS05,SeeBGK07}. It also was found
to be useful for numerical analysis of correlation functions \cite{CauHM05,PerSCHMWA06,PerSCHMWA07,CauCS07}.

The first
result concerning the scalar products in the models with $\mathfrak{gl}(3)$-invariant
$R$-matrix was obtained  in \cite{Res86} and led to the so-called Reshetikhin formula. There, an analog of Izergin--Korepin
formula  for the scalar product of generic Bethe vectors  and a determinant representation
for the norm of the on-shell vectors were found. Similar results for the models based on a $q$-deformed $\mathfrak{gl}(3)$
 were obtained in \cite{PakRS14a,Sla15a}. Recently various particular cases of scalar products
were studied in \cite{BelPR,EscGSVa,EscGSVb,Whe12,PozOK12,Whe13}. However, up to now an analog of the determinant for the scalar product between
generic and on-shell Bethe vectors is not known.  Thus, the $\mathfrak{gl}(3)$ case appears to be more restrictive in this sense.
Apparently, in order to obtain a compact determinant formula for the scalar product in the $\mathfrak{gl}(3)$-based models one should
impose certain constraints on the Bethe parameters of both vectors. In particular, a determinant representation for the scalar product of on-shell and twisted
on-shell Bethe vectors\footnote{See section~\ref{S-N} for the definition of twisted on-shell Bethe vectors.} was derived in \cite{BelPRS12b}.
Later, determinant presentations for form factors of the monodromy matrix entries were obtained in the series of papers \cite{BelPRS13a,PakRS14b,PakRS15a,PakRS15c}.

In the case of  models with $\mathfrak{gl}(2|1)$ symmetry we deal with an intermediate situation. Currently a determinant formula for the scalar product
of generic and on-shell Bethe vectors is not known. However, we succeeded to find a determinant representation for the scalar product in the case where
certain constraints are imposed for the Bethe parameters of both vectors. These constraints are much less restrictive than in the case of
the $\mathfrak{gl}(3)$-based models. Actually, we are able to derive the determinant representation when only a part of Bethe equations for the Bethe parameters
is valid. Thus, a part of Bethe parameters of both vectors is fixed, while the remaining parameters still are generic complex numbers. We call such vectors {\it semi-on-shell} Bethe vectors. The determinant representation for the scalar product of semi-on-shell Bethe vectors in the models with $\mathfrak{gl}(2|1)$ symmetry is the main result of the present paper.

The article is organized as follows. In section~\ref{S-N} we introduce the model under consideration. There we
also specify our conventions and notation. Section~\ref{S-MR} contains the main results of the paper. There we give a determinant
representation for a special  scalar product of  Bethe vectors and consider important particular cases.
In section~\ref{S-O} we prove orthogonality of on-shell Bethe vectors.
In section~\ref{S-FFDE} we  calculate form factors of diagonal entries of the monodromy matrix.
Finally in  section~\ref{S-CSP} we give the derivation of the determinant representation for the scalar product.
Appendix~\ref{A-CI} contains  several auxiliary lemmas allowing us to calculate certain multiple sums of rational functions.

\section{Description of the model\label{S-N}}

\subsection{$\mathfrak{gl}(2|1)$-based models}

The $R$-matrix of $\mathfrak{gl}(2|1)$-based models acts in the tensor product $\mathbb{C}^{2|1}\otimes \mathbb{C}^{2|1}$,
where $\mathbb{C}^{2|1}$ is the $\mathbb{Z}_2$-graded vector space with the grading $[1]=[2]=0$, $[3]=1$.
 Matrices acting in this space are also graded, according to $[e_{ij}]=[i]+[j]$, where $e_{ij}$ are elementary units: $(e_{ij})_{ab}=\delta_{ia}\delta_{jb}$.
The $R$-matrix has the form
 \be{R-mat}
 R(u,v)=\mathbb{I}+g(u,v)P, \qquad g(u,v)=\frac{c}{u-v},
 \ee
where $\mathbb{I}$ is the identity matrix, $P$ is the graded permutation operator \cite{KulS80}, and $c$ is a constant.

 The elements of the monodromy matrix $T(u)$ are graded in the same way as the matrices $[e_{ij}]$: $[T_{ij}(u)]=[i]+[j]$. Their commutation relations
are given by the $RTT$-relation
\be{RTT}
R(u,v)\bigl(T(u)\otimes \mathbb{I}\bigr) \bigl(\mathbb{I}\otimes T(v)\bigr)= \bigl(\mathbb{I}\otimes T(v)\bigr)
\bigl( T(u)\otimes \mathbb{I}\bigr)R(u,v),
\ee
where the tensor products
of $\mathbb{C}^{2|1}$ spaces are  graded as follows:
\be{tens-prod}
(\mathbb{I}\otimes e_{ij})\,\cdot\,(e_{kl}\otimes \mathbb{I}) = (-1)^{([i]+[j])([k]+[l])}\,e_{kl}\otimes e_{ij}.
\ee
Equation \eqref{RTT} holds in the tensor product $\mathbb{C}^{2|1}\otimes \mathbb{C}^{2|1}\otimes\mathcal{H}$,
where $\mathcal{H}$ is a Hilbert space of the Hamiltonian  under consideration.

The $RTT$-relation \eqref{RTT} implies a set of scalar commutation relations for the monodromy matrix elements
\begin{equation}\label{TM-1}
\begin{split}
[T_{ij}(u),T_{kl}(v)\}&
=(-1)^{[i]([k]+[l])+[k][l]}g(u,v)\Big(T_{kj}(v)T_{il}(u)-T_{kj}(u)T_{il}(v)\Big),\\
&=(-1)^{[l]([i]+[j])+[i][j]}g(u,v)\Big(T_{il}(u)T_{kj}(v)-T_{il}(v)T_{kj}(u)\Big),
\end{split}
\end{equation}
where we introduced the graded commutator
\be{Def-SupC}
[T_{ij}(u),T_{kl}(v)\}= T_{ij}(u)T_{kl}(v) -(-1)^{([i]+[j])([k]+[l])}   T_{kl}(v)  T_{ij}(u).
\ee

The graded transfer matrix is defined as the supertrace of the monodromy matrix
\be{transfer.mat}
\mathcal{T}(u)=\str T(u)= \sum_{j=1}^{3} (-1)^{[j]}\, T_{jj}(u).
\ee
It defines an integrable system, due to the relation $[\mathcal{T}(u)\,,\,\mathcal{T}(v)]=0$.

\subsection{Bethe vectors}

Bethe vectors belong to the space in which the Hamiltonian of the model acts. They can be constructed as
certain polynomials in the operators $T_{ij}$ with $i<j$ applied to the pseudovacuum vector $|0\rangle$.
In this paper we do not use an explicit form of these polynomials, however, the reader can find it
in \cite{PakRS16a,HutLPRS16a}. Dual Bethe vectors belong to the dual space, and they are
polynomials in $T_{ij}$ with $i>j$ applied from the right to the dual pseudovacuum vector $\langle0|$.

We denote Bethe vectors and their dual respectively by $\mathbb{B}_{a,b}(\bu;\bv)$ and $\mathbb{C}_{a,b}(\bu;\bv)$. They are parameterized by two sets of
complex parameters (Bethe parameters) $\bu=\{u_1,\dots,u_a\}$ and $\bv=\{v_1,\dots,v_b\}$ with $a,b=0,1,\dots$.
If the Bethe parameters are generic complex numbers, then we say that the corresponding (dual) Bethe vector is generic. If both sets
$\bu$ and $\bv$ are empty, then $\mathbb{B}_{0,0}(\emptyset;\emptyset)=|0\rangle$ and $\mathbb{C}_{0,0}(\emptyset;\emptyset)=\langle0|$.
The vectors $|0\rangle$ and $\langle0|$
are singular vectors for the entries of the monodromy matrix
 \be{Tjj}
 \begin{aligned}
 &T_{ii}(u)|0\rangle=\lambda_i(u)|0\rangle, \qquad   \langle0|T_{ii}(u)=\lambda_i(u)\langle0|,\qquad i=1,2,3,\\
 & T_{ji}(u)|0\rangle=0, \qquad\qquad\quad   \langle0|T_{ij}(u)=0\,,\qquad\qquad 1\leq i<j\leq3\,,
 \end{aligned}
 \ee
where $\lambda_i(u)$ are some scalar functions. In the framework of the generalized model \cite{Kor82} considered in this paper, they remain free functional parameters.
Below it will be convenient to deal with ratios of these functions
 \be{ratios}
 r_1(u)=\frac{\lambda_1(u)}{\lambda_2(u)}, \qquad  r_3(u)=\frac{\lambda_3(u)}{\lambda_2(u)}.
 \ee

A (dual) Bethe vector becomes an eigenvector of the transfer matrix $\mathcal{T}(w)$, if the Bethe parameters satisfy a system
of Bethe equations
\be{AEigenS-1}
\begin{aligned}
r_1(u_j)&=\prod_{\substack{k=1\\k\ne j}}^a\frac{f(u_j,u_k)}{f(u_k,u_j)}\prod_{l=1}^bf(v_l,u_j),\qquad j=1,\dots,a,\\
r_3(v_j)&=\prod_{l=1}^af(v_j,u_l),\qquad j=1,\dots,b,
\end{aligned}
\ee
where we introduced the function $f(u,v)$
\be{f}
f(u,v)=1+g(u,v)=\frac{u-v+c}{u-v}.
\ee
We call such  vectors on-shell Bethe vectors (or dual on-shell Bethe vectors). One has for them
\be{Left-act}
\mathcal{T}(w)\mathbb{B}_{a,b}(\bu;\bv)= \tau(w|\bu,\bv)\,\mathbb{B}_{a,b}(\bu;\bv),\qquad
 \mathbb{C}_{a,b}(\bu;\bv)\mathcal{T}(w) = \tau(w|\bu,\bv)\,\mathbb{C}_{a,b}(\bu;\bv),
\ee
where
\be{tau-def}
\tau(w|\bu,\bv)=\lambda_1(w)\prod_{j=1}^af(u_j,w)+ \lambda_2(w)\prod_{j=1}^af(w,u_j)\prod_{k=1}^b f(v_k,w)
 -\lambda_3(w)\prod_{k=1}^bf(v_k,w).
\ee

Apart from the usual monodromy matrix it is convenient to consider a twisted monodromy matrix $T_{\kappa}(u)$
\cite{IzeK84,KitMST02,BelPRS12b,BelPRS13a}.
For the models with $\mathfrak{gl}(2|1)$ symmetry it is defined as follows. Let $\kappa$ be a $3\times 3$ diagonal matrix
$\kappa=\diag(\kappa_1,\kappa_2,\kappa_3)$, where $\kappa_i$ are complex numbers. Then $T_{\kappa}(u)=\kappa T(u)$,
where $T(u)$ is the standard monodromy matrix.

One can easily check that the twisted monodromy matrix satisfies the $RTT$-relation \eqref{RTT} with the $R$-matrix \eqref{R-mat}.
The supertrace of  the twisted monodromy matrix $\mathcal{T}_{\kappa}(u)=\str T_{\kappa}(u)$ is called the twisted transfer matrix. The eigenstates (resp. dual eigenstates) of
the twisted transfer matrix are called twisted on-shell Bethe vectors (resp. twisted dual on-shell Bethe vectors). A generic
(dual) Bethe vector becomes a twisted (dual) on-shell Bethe vector, if the Bethe parameters satisfy a system
of twisted Bethe equations
\be{ATEigenS-1}
\begin{aligned}
r_1(u_j)&=\frac{\kappa_2}{\kappa_1}\prod_{\substack{k=1\\k\ne j}}^a\frac{f(u_j,u_k)}{f(u_k,u_j)}\prod_{l=1}^bf(v_l,u_j),\qquad j=1,\dots,a,\\
r_3(v_j)&=\frac{\kappa_2}{\kappa_3}\prod_{l=1}^af(v_j,u_l),\qquad j=1,\dots,b.
\end{aligned}
\ee
Then
\be{TLeft-act}
\mathcal{T}_{\kappa}(w)\mathbb{B}_{a,b}(\bu;\bv)= \tau_{\kappa}(w|\bu,\bv)\,\mathbb{B}_{a,b}(\bu;\bv),\qquad
 \mathbb{C}_{a,b}(\bu;\bv)\mathcal{T}_{\kappa}(w) = \tau_{\kappa}(w|\bu,\bv)\,\mathbb{C}_{a,b}(\bu;\bv),
\ee
where the eigenvalue $\tau_{\kappa}(w|\bu,\bv)$ is given by \eqref{tau-def}, in which one should replace $\lambda_i(w)$
by $\kappa_i\lambda_i(w)$ (see also \eqref{Ttau-def}).

In the framework of the generalized model one can consider  the Bethe parameters $\{\bu,\bv\}$ and the functions
$\{r_1(u_j),r_3(v_k)\}$ as two types of variables \cite{Kor82}. The first type comes from the $R$-matrix, the second type comes
from the monodromy matrix. In the case of generic Bethe vectors these two types of variables are independent. However,
in the case of (twisted) on-shell Bethe vectors they become related by the (twisted) Bethe equations.

One can also consider an intermediate case, when only a  subset of  $\{r_1(u_j),r_3(v_k)\}$ is related to the Bethe
parameters $\{\bu,\bv\}$ by a part of the Bethe equations.
For instance, we can impose the first set of equations \eqref{AEigenS-1} involving
the functions $r_1(u_j)$, without imposing the second set of equations for $r_3(v_j)$ (or vice versa). In the case of a concrete model
this means that a part of the Bethe parameters remains free, while the other parameters become functions of them. We call a Bethe
vector possessing this property a {\it semi-on-shell Bethe vector}. Thus, the semi-on-shell Bethe vectors occupy an intermediate position between
generic and on-shell Bethe vectors.  Below we will consider the scalar products of these vectors.

\subsection{Notation}

In this paper we use the same notation and conventions as in \cite{HutLPRS16c}.  Let us recall them.

Besides the functions $g(u,v)$ and $f(u,v)$
described above we introduce also two functions
\be{desand}
\begin{aligned}
h(u,v)&=\frac{f(u,v)}{g(u,v)}=\frac{u-v+c}{c},\\
t(u,v)&=\frac{g(u,v)}{h(u,v)}=g(u,v)-\frac{1}{h(u,v)}=\frac{c^2}{(u-v)(u-v+c)}.
\end{aligned}
\ee
These functions possess the following obvious properties:
 \be{propert}
 g(u,v)=-g(v,u),\quad h(u,v+c)=\frac1{g(u,v)},\quad  f(u,v+c)=\frac1{f(v,u)},\quad  t(u,v+c)=t(v,u).
 \ee

Let us formulate now a convention on the notation.
We  denote sets of variables by bar: $\bx$, $\bu$, $\bv$ etc.
Individual elements of the sets are denoted by latin subscripts: $v_j$, $u_k$ etc. As a rule, the number of elements in the
sets is not shown explicitly in the equations, however we give these cardinalities in
special comments to the formulas. The notation $\bu\pm c$ means that all the elements of the set $\bu$ are
shifted by $\pm c$:  $\bu\pm c=\{u_1\pm c,\dots,u_n\pm c\}$. A union of sets is denoted by braces: $\{\bu,\bv\}\equiv
\bu\cup \bv$.

Subsets of variables are labeled by roman subscripts: $\bu_{\so}$, $\bv_{\rm ii}$, $\bx_{\st}$ etc.
A notation $\bu\Rightarrow\{\bu_{\so},\bu_{\st}\}$ means that the
set $\bu$ is divided into two  subsets $\bu_{\so}$ and $\bu_{\st}$ such that $\{\bu_{\so},\bu_{\st}\}=\bu$ and  $\bu_{\so}\cap\bu_{\st}
=\emptyset$. We assume that the elements in every subset of variables are ordered in such a way that the sequence of
their subscripts is strictly increasing. We call this ordering the natural order.

In order to avoid too cumbersome formulas we use a shorthand notation for products of  functions depending on one or two variables.
 Namely, if the functions $r_k$ \eqref{ratios} or the functions $g$, $f$, $h$ depend
on a set of variables, this means that one should take the product over the corresponding set.
For example,
 \be{SH-prodllll}
 r_1(\bu)= \prod_{u_j\in\bu} r_1(u_j);\quad
 f(u_k,\bv)=\prod_{v_l\in\bv} f(u_k,v_l);\quad
  g(\bv_{\so},\bv_{\st})=\prod_{v_j\in\bv_{\so}}\prod_{v_k\in\bv_{\st}} g(v_j,v_k).
 \ee

Being written in the shorthand notation the eigenvalue of the twisted transfer matrix takes the form
\be{Ttau-def}
\tau_{\kappa}(w|\bu,\bv)=\kappa_1\lambda_1(w)f(\bu,w)+ \kappa_2\lambda_2(w)f(w,\bu) f(\bv,w)
 -\kappa_3\lambda_3(w)f(\bv,w).
\ee
We draw attention of the reader that the eigenvalue $\tau_{\kappa}(w|\bu,\bv)$ depends on the sets $\bu$ and $\bv$ by definition,
and thus, this is not a function of one or two variables. Hence, our shorthand notation does not apply to
$\tau_{\kappa}(w|\bu,\bv)$, and one does not have any product over the sets $\bu$ and $\bv$ in the l.h.s.
of \eqref{Ttau-def}.  On the contrary, in the r.h.s. of \eqref{Ttau-def} we do have  products of $f$-functions over the sets $\bu$ and $\bv$.

Another example of a function depending on two sets of variables
is the partition function of the six-vertex model with domain wall boundary conditions (DWPF) \cite{Kor82,Ize87}.  We denote it by
$K_n(\bu|\bv)$. It depends on two sets of variables $\bu$ and $\bv$, the subscript indicates that
$\#\bu=\#\bv=n$. The function $K_n$ has the following determinant representation
\begin{equation}\label{K-def}
K_n(\bu|\bv)
=\Delta'_n(\bu)\Delta_n(\bv)h(\bu,\bv)
\det_n t(u_j,v_k),
\end{equation}
where $\Delta'_n(\bu)$ and $\Delta_n(\bv)$ are defined by
\be{def-Del}
\Delta'_n(\bu)
=\prod_{j<k}^n g(u_j,u_k),\qquad {\Delta}_n(\bv)=\prod_{j>k}^n g(v_j,v_k).
\ee
It is easy to see that $K_n$ is symmetric over $\bu$ and symmetric over $\bv$, however  $K_n(\bu|\bv)\ne
 K_n(\bv|\bu)$.

\subsection{Scalar products of Bethe vectors}

The scalar product of Bethe vectors is defined as
\be{Def-SP}
S_{a,b}\equiv S_{a,b}(\buc;\bvc|\bub;\bvb)=\mathbb{C}_{a,b}(\buc;\bvc)\mathbb{B}_{a,b}(\bub;\bvb)\,,
\ee
where all the Bethe parameters are {\it a priori} generic complex numbers. We have added the superscripts $C$ and $B$
to the sets $\bu$, $\bv$ in order to stress that the vectors
$\mathbb{C}_{a,b}$ and $\mathbb{B}_{a,b}$ may depend on different sets of parameters.

The main result of paper \cite{HutLPRS16c} is a sum formula for the scalar product of generic Bethe vectors. It gives an explicit
representation for $S_{a,b}$ as a sum over partitions of the Bethe parameters
\begin{multline}\label{Sab-start}
S_{a,b}=\sum  r_1(\buc_{\st})r_1(\bub_{\so})r_3(\bvc_{\st})r_3(\bvb_{\so})f(\buc_{\so},\buc_{\st})f(\bub_{\st},\bub_{\so})g(\bvc_{\so},\bvc_{\st})g(\bvb_{\st},\bvb_{\so})\\
\times \frac{f(\bvc_{\so},\buc_{\so})f(\bvb_{\st},\bub_{\st})}{ f(\bvc,\buc)f(\bvb,\bub)}\;
  Z_{a-k,n}(\buc_{\st},\bub_{\st}|\bvc_{\so},\bvb_{\so})\;Z_{k,b-n}(\bub_{\so},\buc_{\so}|\bvb_{\st},\bvc_{\st}).
\end{multline}
Here the sum is taken over the partitions
 \be{part-1}
 \begin{array}{ll}
 \buc\Rightarrow\{\buc_{\so},\;\buc_{\st}\}, &\qquad  \bvc\Rightarrow\{\bvc_{\so},\;\bvc_{\st}\},\\
 \bub\Rightarrow\{\bub_{\so},\;\bub_{\st}\}, &\qquad  \bvb\Rightarrow\{\bvb_{\so},\;\bvb_{\st}\} .
 \end{array}
 \ee
The partitions are independent except that $\#\bub_{\so}=\#\buc_{\so}=k$ with $k=0,\dots,a$, and $\#\bvb_{\so}=\#\bvc_{\so}=n$
with $n=0,\dots,b$.

The functions $Z_{a-k,n}(\buc_{\st},\bub_{\st}|\bvc_{\so},\bvb_{\so})$ and $Z_{k,b-n}(\bub_{\so},\buc_{\so}|\bvb_{\st},\bvc_{\st})$ are the
highest coefficients \cite{HutLPRS16c}. By definition, they depend on sets of variables, therefore the convention on the shorthand notation is not
applied to them. On the contrary, we use this convention for the functions $r_k$, $f$, and $g$. Thus, in \eqref{Sab-start} every such function should
be understood as the product over the corresponding subset.

In the $\mathfrak{gl}(2|1)$ case the highest coefficient has a determinant representation \cite{HutLPRS16c}, however, we do not use it. Instead, we use two representations in terms of  sums over partitions. The first representation reads
\begin{equation}\label{Zl-part}
Z_{a,b}(\bt;\bx|\bs;\by)=
\sum g(\bom_{\st},\bom_{\so}) h(\bom_{\st},\bx)
g(\bom_{\st},\by)\; K_a(\bom_{\so}|\bt),
\end{equation}
where $K_a(\bom_{\so}|\bt)$ is the DWPF \eqref{K-def}.
The sum is taken over partitions $\{\bx,\bs\}=\bom\Rightarrow\{\bom_{\so},\bom_{\st}\}$ satisfying restrictions
$\#\bom_{\so}=a$, $\#\bom_{\st}=b$. The second representation has the following form:
\begin{equation}\label{ALt-sum}
Z_{a,b}(\bt;\bx|\bs;\by)=f(\bs,\bt)f(\by,\bx) \sum g(\bet_{\so},\bet_{\st}) \frac{h(\bt,\bet_{\st})}{h(\bs,\bet_{\st})}
K_a(\bx|\bet_{\so}).
\end{equation}
Here the sum is taken over partitions $\{\bt,\by+c\}=\bet\Rightarrow\{\bet_{\so},\bet_{\st}\}$ satisfying restrictions
$\#\bet_{\so}=a$, $\#\bet_{\st}=b$.

Let us stress once more that equation \eqref{Sab-start} describes the scalar product in the most general case, when all the
Bethe parameters are generic complex numbers. If Bethe parameters obey certain additional restrictions, then the representation
\eqref{Sab-start} can be simplified. In particular, such a simplification occurs for  semi-on-shell Bethe vectors.

\section{Main results\label{S-MR}}

\subsection{Scalar product of semi-on-shell Bethe vectors\label{S-R1}}

The main result of this paper is a determinant representation for the scalar products of semi-on-shell Bethe vectors.
Consider the scalar product \eqref{Def-SP},
where the following constraints are imposed:
\be{constr}
\begin{aligned}
r_1(\uc_j)&=\varkappa\prod_{\substack{k=1\\k\ne j}}^a\frac{f(\uc_j,\uc_k)}{f(\uc_k,\uc_j)}\prod_{l=1}^bf(\vc_l,\uc_j),\qquad &j=1,\dots,a,\\
r_3(\vb_k)&=\prod_{l=1}^af(\vb_k,\ub_l),\qquad &k=1,\dots,b.
\end{aligned}
\ee
Here $\varkappa$ is a complex parameter. If we set $\varkappa=\kappa_2/\kappa_1$, then we easily recognize the first set of equations
\eqref{ATEigenS-1} for the parameters $\{\buc,\bvc\}$ and the second set of equations \eqref{AEigenS-1}
for the parameters $\{\bub,\bvb\}$. Thus, $\mathbb{B}_{a,b}(\bub;\bvb)$ is a semi-on-shell Bethe vector, while $\mathbb{C}_{a,b}(\buc;\bvc)$
is a dual twisted semi-on-shell Bethe vector.

Let $\bx=\{\bub,\bvc\}$. Define an $(a+b)\times(a+b)$ matrix $\mathcal{N}$ with the following entries:
 \be{N1}
 \mathcal{N}_{jk}=\frac{(-1)^{a-1}r_1(x_k)}{f(\bvc,x_k)}
 t(\uc_j,x_k)h(\buc,x_k) +\varkappa
  t(x_k,\uc_j)h(x_k,\buc), \qquad \begin{array}{l}j=1,\dots,a\\ k=1,\dots,a+b,\end{array}
 \ee
and
 \begin{multline}\label{N2}
 \mathcal{N}_{a+j,k}= \frac{g(x_k,\bvb)}{g(x_k,\bvc)}\left(1-\frac{r_3(x_k)}{f(x_k,\bub)}\right)\\
 \times \left(g(x_{k},\vc_{j})h(x_k,\bub) +\frac{(-1)^{a-1}r_1(x_{k})r_3(\vc_{j})h(\bub,x_k)}
 {\varkappa f(\vc_j,\buc)f(\bvb,x_k)h(\vc_{j},x_{k})}\right),\qquad \begin{array}{l}j=1,\dots,b,\\ k=1,\dots,a+b.\end{array}
 \end{multline}

\begin{prop}\label{prop:main}
The scalar product $S_{a,b}$ \eqref{Def-SP} with constraint  \eqref{constr} has the following determinant representation
\be{result}
S_{a,b}=\Delta_{a+b}(\bx)\Delta'_a(\buc)\Delta'_b(\bvc)\det_{a+b}\mathcal{N},
\ee
where $\Delta$ and $\Delta'$ are defined in \eqref{def-Del}.
\end{prop}
The proof of representation \eqref{result} will be given in section~\ref{S-CSP}.

It follows from \eqref{Sab-start} that the scalar products are symmetric under the simultaneous replacement
$\buc\leftrightarrow\bub$ and $\bvc\leftrightarrow\bvb$. Therefore making this
replacement  in \eqref{constr}--\eqref{result}  we obtain a
determinant presentation for the scalar product of another set of semi-on-shell Bethe vectors.

It is interesting to see how the matrix elements $\mathcal{N}_{jk}$ depend on the functions $r_1$ and $r_3$. Namely, one
can easily  show that $\mathcal{N}_{jk}$ might depend on $r_3(\vc_k)$, however, they do not depend on $r_3(\ub_k)$. Indeed,
if $x_k=\ub_k$ in \eqref{N2}, then $r_3(x_k)$ is multiplied with the product $1/f(x_k,\bub)$, which vanishes for $x_k\in\bub$.
Similarly, the function $r_1(x_k)$ always enters either with the product $1/f(\bvc,x_k)$ or $1/g(\bvc,x_k)$. Both these products
vanish for $x_k\in\bvc$, therefore the matrix elements $\mathcal{N}_{jk}$ might depend on $r_1(\ub_k)$ only.

Finally, looking at \eqref{N2} for $k>a$ (that is, $x_k\in\bvc$), we see that $\mathcal{N}_{jk}\sim \delta_{jk}$ due to the
product $1/g(\bvc,x_k)$. Thus, the right-lower block of the matrix $\mathcal{N}$ is diagonal.

\subsection{Scalar product of twisted and usual on-shell Bethe vectors\label{S-R2}}

Equation \eqref{result} has important particular cases. First of all, it describes a scalar product of twisted and
usual on-shell Bethe vectors. Let the twist matrix be $\kappa=\diag(\kappa_1,\kappa_2,\kappa_3)$. Then one should set
$\varkappa=\kappa_2/\kappa_1$ in \eqref{constr} and impose two additional constraints
\be{add-constr}
\begin{aligned}
r_1(\ub_j)&=\prod_{\substack{k=1\\k\ne j}}^a\frac{f(\ub_j,\ub_k)}{f(\ub_k,\ub_j)}\prod_{l=1}^bf(\vb_l,\ub_j), \qquad &j=1,\dots,a,\\
r_3(\vc_k)&=\frac{\kappa_2}{\kappa_3}\prod_{l=1}^af(\vc_k,\uc_l),\qquad &k=1,\dots,b.
\end{aligned}
\ee
In this case the vector $\mathbb{B}_{a,b}(\bub;\bvb)$ becomes an on-shell Bethe vector, while $\mathbb{C}_{a,b}(\buc;\bvc)$ becomes
a twisted on-shell Bethe vector. Then their scalar product has the determinant representation \eqref{result}, where
 \be{N1-k}
 \mathcal{N}_{jk}=h(x_k,\bub)\left(t(\uc_j,x_k)\frac{f(\bvb,x_k)h(\buc,x_k) }{f(\bvc,x_k)h(\bub,x_k)} +\frac{\kappa_2}{\kappa_1}
  t(x_k,\uc_j)\frac{h(x_k,\buc)}{h(x_k,\bub)}\right), \quad j=1,\dots,a,
 \ee
 \begin{equation}\label{N2-k}
\mathcal{N}_{a+j,k} = h(x_k,\bub)\frac{g(x_k,\bvb)}{g(x_k,\bvc)}\left(1-\frac{\kappa_2}{\kappa_3}\frac{f(x_k,\buc)}{f(x_k,\bub)}\right)
  \left(g(x_{k},\vc_{j}) +\frac{\kappa_1} {\kappa_3 h(\vc_{j},x_{k})}\right),\quad
 j=1,\dots,b,
 \end{equation}
and $k=1,\dots,a+b$ in both formulas.

\subsection{Norm of on-shell Bethe vector\label{S-R3}}

The second particular case of \eqref{result} is the norm of on-shell Bethe vector. For this one should set\footnote{%
Here and below the notation $\bar\kappa=1$ means $\kappa_1=\kappa_2=\kappa_3=1$.}
$\bar\kappa=1$ and consider the limit $\ub_j\to\uc_j=u_j$, $\vc_j\to\vb_j=v_j$.
Then the result has the form
\be{Norm-res}
\|\mathbb{B}_{a,b}(\bu;\bv)\|^2=(-1)^{a+b}\prod_{j=1}^b\prod_{k=1}^a f(v_j,u_k)\prod_{\substack{j,k=1\\j\ne k}}^a f(u_j,u_k)\;
\prod_{\substack{j,k=1\\j\ne k}}^b g(v_j,v_k)\;\det_{a+b}\widehat{\mathcal{N}}.
\ee
Here $\widehat{\mathcal{N}}$ is an $(a+b)\times(a+b)$ block-matrix. The left-upper block is
 \begin{equation}\label{P11-n2}
\widehat{\mathcal{N}}_{jk}=
 \delta_{jk}\left[c\,\frac{r'_1(u_k)}{r_1(u_k)}+\sum_{\ell=1}^a\frac{2c^2}{u_{k\ell}^2-c^2}-\sum_{m=1}^b t(v_m,u_k)  \right]-
 \frac{2c^2}{u_{kj}^2-c^2},\quad j,k=1,\dots,a,
 \end{equation}
where $u_{kj}=u_k-u_j$ and $r'_1(u_k)$ means the derivative of the function $r_1(u)$ at the point $u=u_k$. The right-lower block is
diagonal
 \begin{equation}\label{P22-n2}
\widehat{\mathcal{N}}_{j+a,k+a}=\delta_{jk}
 \left[c\,\frac{r'_3(v_k)}{r_3(v_k)}+\sum_{\ell=1}^a t(v_k,u_\ell)  \right],\quad j,k=1,\dots,b,
 \end{equation}
where $r'_3(v_k)$ means the derivative of the function $r_3(v)$ at the point $v=v_k$.
The antidiagonal blocks are
 \begin{equation}\label{P12-n2}
\widehat{\mathcal{N}}_{j,k+a}=t(v_k,u_j), \quad j=1,\dots,a,\quad k=1,\dots,b,
 \end{equation}
 and
 \begin{equation}\label{P21-n2}
\widehat{\mathcal{N}}_{j+a,k}=-t(v_j,u_k),\quad j=1,\dots,b,\quad k=1,\dots,a.
 \end{equation}

It is easy to relate the determinant of the matrix $\widehat{\mathcal{N}}$ with the Jacobian of the
Bethe equations. Namely, let
\be{Phi}
\begin{aligned}
&\Phi_j= \log\Big(\frac{r_1(u_j)}{f(\bv,u_j)}\prod_{\substack{k=1\\k\ne j}}^a\frac{f(u_k,u_j)}{f(u_j,u_k)}\Big),\qquad j=1,\dots,a,\\
&\Phi_{a+j}=\log\left(\frac{r_3(v_j)}{f(v_j,\bu)}\right),\qquad j=1,\dots,b.
\end{aligned}
\ee
Then the Bethe equations for the sets $\bu$ and $\bv$ take the form
\be{BE-log}
\Phi_j=2\pi in_j, \qquad j=1,\dots,a+b,
\ee
where $n_j$ are integer numbers. A straightforward calculation shows that
\be{Norm-Phi}
\begin{aligned}
&\widehat{\mathcal{N}}_{j,k}=c\,\frac{\partial\Phi_j}{\partial u_k}
\qquad &k=1,\dots,a,\\
&\widehat{\mathcal{N}}_{j,a+k}=c\,\frac{\partial\Phi_j}{\partial v_k},\qquad &k=1,\dots,b,
\end{aligned}
\qquad\qquad
 j=1,...,a+b.
\ee

\section{Orthogonality of the eigenvectors \label{S-O}}

Consider the scalar product of twisted on-shell and usual on-shell vectors. In this case the entries of the matrix $\mathcal{N}$ are
given by \eqref{N1-k}, \eqref{N2-k}. Assume
that $\{\buc,\bvc\}\ne \{\bub,\bvb\}$ at $\kappa = 1$. Then in the limit $\kappa = 1$ we obtain the scalar product of two different
on-shell Bethe vectors, which should be orthogonal. Let us show this.

To prove the orthogonality of on-shell Bethe vectors we introduce an $(a+b)$-component vector $\Omega$
\be{vector}
\begin{aligned}
\Omega_j&=\frac{1}{g(\uc_j,\bub)}\prod_{\substack{k=1\\k\ne j}}^ag(\uc_j,\uc_k),\qquad j=1,\dots,a,\\
\Omega_{a+j}&=\frac{1}{g(\vc_j,\bvb)}\prod_{\substack{k=1\\k\ne j}}^ag(\vc_j,\vc_k),\qquad j=1,\dots,b.
\end{aligned}
\ee
Due to the condition $\{\buc,\bvc\}\ne \{\bub,\bvb\}$, the vector $\Omega$ has at least one non-zero component.

Using the contour integral method (see Appendix~\ref{A-CI1}) one can easily calculate the following sums:
\be{sum-rows}
\begin{aligned}
&\sum_{j=1}^at(\uc_j,x_k)\Omega_j=\frac{h(\bub,x_k)}{h(\buc,x_k)}-\frac{g(x_k,\buc)}{g(x_k,\bub)},
\qquad\quad
\sum_{j=1}^b g(x_k,\vc_j)\Omega_{a+j}=\frac{g(x_k,\bvc)}{g(x_k,\bvb)}-1,\\
&\sum_{j=1}^at(x_k,\uc_j)\Omega_j=\frac{g(x_k,\buc)}{g(x_k,\bub)}-\frac{h(x_k,\bub)}{h(x_k,\buc)},
\qquad\quad
\sum_{j=1}^b \frac{\Omega_{a+j}}{h(\vc_j,x_k)}=1-\frac{h(\bvb,x_k)}{h(\bvc,x_k)}.
\end{aligned}
\ee

Using these results we obtain
\begin{equation}\label{sum-N11N21}
\frac{\sum_{j=1}^{a}\mathcal{N}_{jk}\Omega_j}{h(x_k,\bub)}=
  \frac{f(\bvb,x_k)}{f(\bvc,x_k)}\left(1-\frac{f(\buc,x_k)}{f(\bub,x_k)}\right)
  +\frac{\kappa_2}{\kappa_1}\left(\frac{f(x_k,\buc)}{f(x_k,\bub)}-1\right),
\end{equation}
and
\begin{equation}\label{sum-N12N22}
\frac{\sum_{j=1}^{b}\mathcal{N}_{a+j,k}\Omega_{a+j}}{h(x_k,\bub)}=
\left(1-\frac{\kappa_2}{\kappa_3}\frac{f(x_k,\buc)}{f(x_k,\bub)}\right)\left[
\left(\frac{\kappa_1}{\kappa_3}-1\right)\frac{g(\bvb,x_k)}{g(\bvc,x_k)}+1
-\frac{\kappa_1}{\kappa_3}\frac{f(\bvb,x_k)}{f(\bvc,x_k)}\right].
\end{equation}
Note that if $x_k\in\{\bub,\bvc\}$, then $\bigl(f(x_k,\bub)f(\bvc,x_k)\bigr)^{-1}=0$,   $\bigl(f(x_k,\bub)g(\bvc,x_k)\bigr)^{-1}=0$, and $\bigl(f(\bub,x_k)f(\bvc,x_k)\bigr)^{-1}=0$. Therefore the terms proportional to these products
vanish. Thus, neglecting such the terms we find
\begin{equation}\label{sum-Ntot}
\frac{\sum_{j=1}^{a+b}\mathcal{N}_{jk}\Omega_j}{h(x_k,\bub)}=
1-\frac{\kappa_2}{\kappa_1}+
\left(\frac{\kappa_1}{\kappa_3}-1\right)\left(\frac{g(\bvb,x_k)}{g(\bvc,x_k)}-  \frac{f(\bvb,x_k)}{f(\bvc,x_k)}\right)
+\frac{f(x_k,\buc)}{f(x_k,\bub)}\left(\frac{\kappa_2}{\kappa_1}
-\frac{\kappa_2}{\kappa_3}\right).
\end{equation}
We see that this linear combination of rows of the matrix $\mathcal{N}$ vanishes at $\bar\kappa=1$. Hence, the determinant vanishes at
$\bar\kappa=1$, which means that two  different on-shell vectors are orthogonal.

\section{Form factors of diagonal elements\label{S-FFDE}}

We define form factors of the diagonal monodromy matrix entries as matrix elements of the operators $T_{ii}(z)$ between two
on-shell Bethe vectors. We use a standard method for their  calculation \cite{IzeK84,KitMST02,BelPRS12b,BelPRS13a}. Let
\be{Q}
Q=\mathbb{C}^{(\kappa)}_{a,b}(\buc;\bvc)\bigl(\str T_{\kappa}(z)-\str T(z)\bigr)\mathbb{B}_{a,b}(\bub;\bvb).
\ee
Here $T_{\kappa}(z)$ is the twisted monodromy matrix and $T(z)$ is the usual monodromy matrix. We assume that
$\mathbb{B}_{a,b}(\bub;\bvb)$ is an on-shell Bethe vector and $\mathbb{C}^{(\kappa)}_{a,b}(\buc;\bvc)$ is a dual
twisted on-shell Bethe vector. In order to stress this difference we have added a superscript $\kappa$
 on the dual twisted Bethe vector.
Obviously, $\mathbb{C}^{(\kappa)}_{a,b}(\buc;\bvc)$ turns into the usual dual on-shell Bethe vector at $\bar\kappa=1$.

On the one hand
\be{Q1}
Q=\sum_{i=1}^3 (-1)^{[i]}(\kappa_i-1)\mathbb{C}^{(\kappa)}_{a,b}(\buc;\bvc)\; T_{ii}(z)\;\mathbb{B}_{a,b}(\bub;\bvb).
\ee
On the other hand
\be{Q2}
Q=\Big(\tau_{\kappa}(z|\buc,\bvc)-\tau(z|\bub,\bvb)\Big)\mathbb{C}^{(\kappa)}_{a,b}(\buc;\bvc)\mathbb{B}_{a,b}(\bub;\bvb),
\ee
where $\tau_{\kappa}(z|\buc,\bvc)$ and $\tau(z|\bub,\bvb)$ respectively are the eigenvalues of the twisted and usual transfer matrices.
Comparing \eqref{Q1} and \eqref{Q2} we find
\begin{multline}\label{FF-1}
\mathbb{C}_{a,b}(\buc;\bvc)\; T_{ii}(z)\;\mathbb{B}_{a,b}(\bub;\bvb)\\
=(-1)^{[i]}\frac{d}{d\kappa_i}\Bigr[
\bigl(\tau_{\kappa}(z|\buc,\bvc)-\tau(z|\bub,\bvb)\bigr)\mathbb{C}^{(\kappa)}_{a,b}(\buc;\bvc)\mathbb{B}_{a,b}(\bub;\bvb)\Bigr]_{\bar\kappa=1}.
\end{multline}
Here $\mathbb{C}_{a,b}(\buc;\bvc)$  is the value of the vector $\mathbb{C}^{(\kappa)}_{a,b}(\buc;\bvc)$ at $\bar\kappa=1$.
Since this vector is a dual on-shell vector, we obtain a form factor of $T_{ii}(z)$ in the l.h.s. of \eqref{FF-1}.
In the r.h.s. of \eqref{FF-1} we should distinguish between two cases. If
$\{\buc,\bvc\}=\{\bub,\bvb\}=\{\bu,\bv\}$ at $\bar\kappa=1$, then  the derivative in \eqref{FF-1} acts on $\tau_{\kappa}(z|\buc,\bvc)$
only, and we find
\begin{equation}\label{FF-norm}
\mathbb{C}_{a,b}(\bu;\bv)\; T_{ii}(z)\;\mathbb{B}_{a,b}(\bu;\bv)\\
=(-1)^{[i]}\|\mathbb{B}_{a,b}(\bu;\bv)\|^2
\frac{d}{d\kappa_i}\tau_{\kappa}(z|\buc,\bvc)\Bigr|_{\raisebox{-1ex}{$\substack{\buc=\bu, \; \bvc=\bv\\ \bar\kappa=1}$}}\;.
\end{equation}
Note, that here we have the full derivative over $\kappa_i$. Therefore, it acts also on the Bethe parameters $\{\buc,\bvc\}$, because due to
the twisted Bethe equations they implicitly depend on the twist parameters: $\buc=\buc(\kappa)$ and $\bvc=\bvc(\kappa)$.

If $\{\buc,\bvc\}\ne\{\bub,\bvb\}$ at $\bar\kappa=1$, then we define a universal form factor \cite{PakRS15a} of the operator $T_{ii}(z)$ as
\be{FF-univ-def}
\mathfrak{F}_{a,b}^{(ii)}(\buc;\bvc|\bub;\bvb)=\frac{\mathbb{C}_{a,b}(\buc;\bvc)\; T_{ii}(z)\;\mathbb{B}_{a,b}(\bub;\bvb)}
{\tau(z|\buc,\bvc)-\tau(z|\bub,\bvb)}.
\ee
In this case, due to the orthogonality of Bethe vectors at $\{\buc,\bvc\}\ne\{\bub,\bvb\}$,  the derivative in the r.h.s. of \eqref{FF-1} acts on the scalar product only. Therefore we obtain for the universal form factor of $T_{ii}(z)$
\be{FF-2}
\mathfrak{F}_{a,b}^{(ii)}(\buc;\bvc|\bub;\bvb)=(-1)^{[i]}\frac{d}{d\kappa_i}
\mathbb{C}^{(\kappa)}_{a,b}(\buc;\bvc)\mathbb{B}_{a,b}(\bub;\bvb)\Bigr|_{\bar\kappa=1}.
\ee
Thus, for the calculation of the $T_{ii}(z)$ universal form factors,  it is enough to differentiate  w.r.t. $\kappa_i$
the scalar product of the twisted and usual on-shell
vectors and then set $\bar\kappa=1$.

Suppose that  $\Omega_p\ne 0$ for some $p\in\{1,\dots,a+b\}$ in the vector \eqref{vector}.
Then we can add to the $p$-th row of the matrix $\mathcal{N}$ all other rows multiplied with $\Omega_j/\Omega_{p}$.
It is clear that after this the elements of the $p$-th row are given by \eqref{sum-Ntot} with the common prefactor
$\Omega^{-1}_{p}$. Then to calculate the form factor of $T_{ii}(z)$ it is enough to differentiate the $p$-th row with respect to
$\kappa_i$ at $\bar\kappa=1$. Hereby we should simply set $\bar\kappa=1$ everywhere else. We obtain
\begin{equation}\label{FFss-res}
\mathfrak{F}_{a,b}^{(i,i)}(\buc;\bvc|\bub;\bvb)=\Omega^{-1}_{p}\, f(\bvc,\bub)h(\bub,\bub)
\Delta'(\buc)\Delta(\bub)\Delta(\bvc)\Delta'(\bvc)\;
\det_{a+b}\mathcal{N}^{(i)}.
\end{equation}
Here  the entries $\mathcal{N}_{jk}^{(i)}$ of the matrix $\mathcal{N}^{(i)}$  do not depend on $i=1,2,3$, for all $j\ne p$. In other words, they are the same
for all $\mathfrak{F}_{a,b}^{(ii)}$.  More precisely, if $1\le j\le a$, then
 \be{Ni1-k}
 \mathcal{N}^{(i)}_{jk}=t(\uc_j,x_k)\frac{f(\bvb,x_k)h(\buc,x_k) }{f(\bvc,x_k)h(\bub,x_k)} +
  t(x_k,\uc_j)\frac{h(x_k,\buc)}{h(x_k,\bub)}, \qquad k=1,\dots,a+b,
 \ee
and if $1\le j\le b$, then
 \begin{equation}\label{Ni2-k}
 \mathcal{N}^{(i)}_{a+j,k}= -\frac{g(x_k,\bvb)}{g(x_k,\bvc)}\left(1-\frac{f(x_k,\buc)}{f(x_k,\bub)}\right)
 t(\vc_{j},x_{k}),\qquad k=1,\dots,a+b.
 \end{equation}
The $p$-th row depends on the specific universal form factor $\mathfrak{F}_{a,b}^{(ii)}$,  $i=1,2,3$:
 \be{Yk}
 \begin{aligned}
& \mathcal{N}^{(1)}_{pk}=1+\frac{g(\bvb,x_k)}{g(\bvc,x_k)}-  \frac{f(\bvb,x_k)}{f(\bvc,x_k)}-\frac{f(x_k,\buc)}{f(x_k,\bub)}, \\
& \mathcal{N}^{(2)}_{pk}=-1,\\
& \mathcal{N}^{(3)}_{pk}= \mathcal{N}^{(1)}_{pk}+\mathcal{N}^{(2)}_{pk},
 \end{aligned}
 \qquad\qquad k=1,\dots,a+b.
 \ee
Recall   that here $p$ is an arbitrary integer from the set $\{1,\dots,a+b\}$ such that $\Omega_p\ne 0$.
Observe, that equations \eqref{Yk} imply  the vanishing  of the form factors of $\str T(z)$ between different states, in accordance with
the orthogonality of on-shell Bethe vectors.

\section{Calculating the scalar product\label{S-CSP}}

In this section we prove Proposition~\ref{prop:main}. Our starting point is the sum formula \eqref{Sab-start} for the scalar product
of generic Bethe vectors.

\subsection{Summation over the partitions of $\buc$ and $\bvb$\label{S-SOP}}

Substituting  \eqref{Zl-part} for $Z_{a-k,n}(\buc_{\st},\bub_{\st}|\bvc_{\so},\bvb_{\so})$   and \eqref{ALt-sum}
for $Z_{k,b-n}(\bub_{\so},\buc_{\so}|\bvb_{\st},\bvc_{\st})$ into the representation \eqref{Sab-start} we obtain
\begin{multline}\label{Sab-1}
S_{a,b}=
\sum  r_1(\buc_{\st})r_1(\bub_{\so})r_3(\bvc_{\st})r_3(\bvb_{\so})f(\buc_{\so},\buc_{\st})f(\bub_{\st},\bub_{\so})g(\bvc_{\so},\bvc_{\st})g(\bvb_{\st},\bvb_{\so})\\
\times \frac{ K_{a-k}(\bom_{\so}|\buc_{\st})K_k(\buc_{\so}|\bet_{\so})}{ f(\bvc,\buc_{\st})f(\bvb_{\so},\bub)}\;
g(\bom_{\st},\bom_{\so}) h(\bom_{\st},\bub_{\st})g(\bom_{\st},\bvb_{\so})g(\bet_{\so},\bet_{\st}) \frac{h(\bub_{\so},\bet_{\st})}{h(\bvb_{\st},\bet_{\st})}.
\end{multline}
Here we have additional summations over the partitions $\{\bub_{\st},\bvc_{\so}\}=\bom\Rightarrow\{\bom_{\so},\bom_{\st}\}$ and $\{\bub_{\so},\bvc_{\st}+c\}=\bet\Rightarrow\{\bet_{\so},\bet_{\st}\}$, such that $\#\bom_{\so}=a-k$, $\#\bom_{\st}=n$,
$\#\bet_{\so}=k$, and $\#\bet_{\st}=b-n$.

Let suppose that the constraints \eqref{constr} are fulfilled.
Then taking the products of \eqref{constr} with respect to the corresponding subsets we obtain
\be{ATEigenS-m1}
\begin{aligned}
r_1(\buc_{\st})&=\varkappa^{a-k}\frac{f(\buc_{\st},\buc_{\so})}{f(\buc_{\so},\buc_{\st})}f(\bvc,\buc_{\st}),\\
r_3(\bvb_{\so})&=f(\bvb_{\so},\bub).
\end{aligned}
\ee
Substituting these expressions into \eqref{Sab-1} we arrive at
\begin{multline}\label{Sab-2}
S_{a,b}=
\sum \varkappa^{a-k} r_1(\bub_{\so})r_3(\bvc_{\st})f(\bub_{\st},\bub_{\so})g(\bvc_{\so},\bvc_{\st})g(\bom_{\st},\bom_{\so})g(\bet_{\so},\bet_{\st})
\\
\times h(\bub_{\so},\bet_{\st})h(\bom_{\st},\bub_{\st})
 \Bigr[ f(\buc_{\st},\buc_{\so})  K_{a-k}(\bom_{\so}|\buc_{\st})K_k(\buc_{\so}|\bet_{\so})\Bigl]\;
 \cdot\Bigl[ g(\bvb_{\st},\bvb_{\so})
\frac{g(\bom_{\st},\bvb_{\so})}{  h(\bvb_{\st},\bet_{\st}) }\Bigr].
\end{multline}
Here in the second line we have allocated with square brackets the terms depending on the subsets of
$\buc$ and the subsets of $\bvb$  respectively.
The  sum over partitions $\buc\Rightarrow\{\buc_{\so},\;\buc_{\st}\}$ can be computed via \eqref{Sym-Part-old1}:
\be{sum-buc}
\sum K_{a-k}(\bom_{\so}|\buc_{\st})K_k(\buc_{\so}|\bet_{\so})f(\buc_{\st},\buc_{\so})=(-1)^k
f(\buc,\bet_{\so})K_a(\{\bet_{\so}-c,\bom_{\so}\}|\buc).
\ee
The  sum over partitions $\bvb\Rightarrow\{\bvb_{\so},\;\bvb_{\st}\}$ can be computed via
\eqref{Sym-Part-new1}:
\begin{multline}\label{sum-bvb}
\sum g(\bvb_{\st},\bvb_{\so})\frac{g(\bom_{\st},\bvb_{\so})}{  h(\bvb_{\st},\bet_{\st}) }=
(-1)^n\sum g(\bvb_{\st},\bvb_{\so})g(\bvb_{\so},\bom_{\st}) g(\bvb_{\st},\bet_{\st}-c)\\
=(-1)^n\;
\frac{g(\bvb,\bom_{\st})h(\bom_{\st},\bet_{\st})}{h(\bvb,\bet_{\st})}.
\end{multline}
Substituting these results into \eqref{Sab-2} we arrive at
\begin{multline}\label{Sab-3}
S_{a,b}=
\sum \varkappa^{a-k}(-1)^{k+n} r_1(\bub_{\so})r_3(\bvc_{\st})f(\bub_{\st},\bub_{\so})g(\bvc_{\so},\bvc_{\st})
f(\buc,\bet_{\so})\\
\times
g(\bom_{\st},\bom_{\so})g(\bet_{\so},\bet_{\st})g(\bvb,\bom_{\st}) h(\bom_{\st},\bub_{\st})
\frac{h(\bub_{\so},\bet_{\st})h(\bom_{\st},\bet_{\st})}{h(\bvb,\bet_{\st})}\;K_a(\{\bet_{\so}-c,\bom_{\so}\}|\buc).
\end{multline}
 We recall that in this formula
\be{card-1}
\begin{aligned}
&\#\bub_{\so}=k,\qquad &\#\bub_{\st}=a-k,\qquad &\#\bvc_{\so}=n,\qquad &\#\bvc_{\st}=b-n,\\
&\#\bom_{\so}=a-k,\qquad  &\#\bom_{\st}=n,\qquad &\#\bet_{\so}=k,\qquad &\#\bet_{\st}=b-n.
\end{aligned}
\ee

\subsection{Partial summation over the partitions of $\bub$ and $\bvc$\label{S-PSOP}}

In order to go further we specify the subsets in \eqref{Sab-3} as follows
\be{part-2}
 \begin{array}{llll}
 \bub_{\so}=\{\bub_{\rm i},\;\bub_{\rm ii}\},\quad &\bub_{\st}=\{\bub_{\rm iii},\;\bub_{\rm iv}\},\quad &\bom_{\so}=\{\bub_{\rm iii},\;\bvc_{\rm iv}\} ,\quad
& \bom_{\st}=\{\bub_{\rm iv},\;\bvc_{\rm iii}\},
\\
 \bvc_{\so}=\{\bvc_{\rm iii},\;\bvc_{\rm iv}\}, & \bvc_{\st}=\{\bvc_{\rm i},\;\bvc_{\rm ii}\},
& \bet_{\so}=\{\bub_{\rm ii},\;\bvc_{\rm i}+c\},&  \bet_{\st}=\{\bub_{\rm i},\;\bvc_{\rm ii}+c\}.
 \end{array}
 \ee
The cardinalities of the introduced sub-subsets are $\#\bub_j=k_{j}$ and $\#\bvc_j=n_j$
for $j={\rm i},{\rm ii},{\rm iii},{\rm iv}$. It is easy to see that
$k_{\rm ii}+k_{\rm i}=k$, $n_{\rm iii}+n_{\rm iv}=n$, $k_{\rm iv}=n_{\rm iv}$, and $k_{\rm i}=n_{\rm i}$.

Equation \eqref{Sab-3} takes the form
\begin{multline}\label{Sab-4}
S_{a,b}=
\sum \varkappa^{a-k}(-1)^{k+n} r_1(\bub_{\rm i})r_1(\bub_{\rm ii})r_3(\bvc_{\rm i})r_3(\bvc_{\rm ii})
\frac{f(\buc,\bub_{\rm ii})}{ f(\bvc_{\rm i},\buc) }\\
\times
f(\bub_{\rm iii},\bub_{\rm i})f(\bub_{\rm iv},\bub_{\rm i})f(\bub_{\rm iii},\bub_{\rm ii})f(\bub_{\rm iv},\bub_{\rm ii})
g(\bvc_{\rm iii},\bvc_{\rm i})g(\bvc_{\rm iv},\bvc_{\rm i})g(\bvc_{\rm iii},\bvc_{\rm ii})g(\bvc_{\rm iv},\bvc_{\rm ii})
\\
\times
g(\bub_{\rm iv},\bub_{\rm iii})g(\bub_{\rm iv},\bvc_{\rm iv})g(\bvc_{\rm iii},\bub_{\rm iii}) g(\bvc_{\rm iii},\bvc_{\rm iv})
 g(\bub_{\rm ii},\bub_{\rm i})g(\bub_{\rm ii},\bvc_{\rm ii}+c)g(\bvc_{\rm i}+c,\bub_{\rm i})g(\bvc_{\rm i},\bvc_{\rm ii})\\
\times \frac{g(\bvb,\bub_{\rm iv})g(\bvb,\bvc_{\rm iii})}{h(\bvb,\bub_{\rm i})h(\bvb,\bvc_{\rm ii}+c)}
 h(\bub_{\rm iv},\bub_{\rm iii})h(\bub_{\rm iv},\bub_{\rm iv}) h(\bvc_{\rm iii},\bub_{\rm iii})h(\bvc_{\rm iii},\bub_{\rm iv})\\
\times
h(\bub_{\rm i},\bub_{\rm i})h(\bub_{\rm ii},\bub_{\rm i}) h(\bub_{\rm i},\bvc_{\rm ii}+c)h(\bub_{\rm ii},\bvc_{\rm ii}+c)  \\
\times    h(\bub_{\rm iv},\bub_{\rm i}) h(\bub_{\rm iv},\bvc_{\rm ii}+c)h(\bvc_{\rm iii},\bub_{\rm i}) h(\bvc_{\rm iii},\bvc_{\rm ii}+c)
\;K_a(\{\bub_{\rm ii}-c,\bvc_{\rm i},\bvc_{\rm iv},\bub_{\rm iii}\}|\buc).
\end{multline}
Here we have used  the relation $f(x,y+c)=1/f(y,x)$.

Now we combine the sub-subsets of $\bub$ and $\bvc$ into new subsets
 \be{part-3}
 \begin{array}{ll}
 \bub_{\so}=\{\bub_{\rm i},\;\bub_{\rm iv}\},&\qquad \bvc_{\so}=\{\bvc_{\rm i},\;\bvc_{\rm iv}\} ,\\
 \bub_{\st}=\{\bub_{\rm ii},\;\bub_{\rm iii}\},&\qquad  \bvc_{\st}=\{\bvc_{\rm ii},\;\bvc_{\rm iii}\}.
 \end{array}
 \ee
Due to \eqref{part-2} we have $\#\bub_{\so}=\#\bvc_{\so}=n_{\rm i}+n_{\rm iv}\equiv n_{\so}$. Observe that
these new subsets are different from the subsets used, for example, in \eqref{Sab-start}. We use however the
same notation, as we deal with sums over partitions, and therefore it does not matter how we denote the separate terms
  in these sums.

Then using \eqref{propert} we recast \eqref{Sab-4} in a partly factorized form
 \begin{multline}\label{Sab-fact}
S_{a,b}=(-1)^{b}\sum_{\substack{\bub\Rightarrow\{\bub_{\so},\bub_{\st}\}\\ \bvc\Rightarrow \{\bvc_{\so},\;\bvc_{\st}\}}}
  f(\bub_{\so},\bub_{\st})h(\bub_{\so},\bub_{\so})g(\bvc_{\st},\bvc_{\so})  g(\bvb,\bub_{\so})g(\bvb,\bvc_{\st})
  g(\bvc_{\st},\bub_{\st})h(\bvc_{\st},\bub)  \\
  \times
  G_{n_{\so}}(\bub_{\so}|\bvc_{\so}) \mathcal{L}_a^{(u)}(\{\bub_{\st},\bvc_{\so}\}|\buc)
 \mathcal{L}_b^{(v)}(\bvc_{\st}|\bub).
 \end{multline}
Here the  partitions of the sets $\bub$ and $\bvc$ are explicitly shown by the superscripts of the sum.
The functions  $G_{n_{\so}}$, $\mathcal{L}_a^{(u)}$, and $\mathcal{L}_b^{(v)}$ in their turn are given as sums over partitions into sub-subsets.
We have for $G_{n_{\so}}$
\be{G}
G_{n_{\so}}(\bub_{\so}|\bvc_{\so})=
 \sum_{\substack{\bub_{\so}\Rightarrow\{\bub_{\rm iv},\;\bub_{\rm i}\} \\ \bvc_{\so}\Rightarrow\{\bvc_{\rm i},\;\bvc_{\rm iv}\}}}
 \varkappa^{-n_{\rm i}}
\hat r_3(\bvc_{\rm i})\hat r_1(\bub_{\rm i}) g(\bub_{\rm i},\bub_{\rm iv})g(\bvc_{\rm iv},\bvc_{\rm i})\frac{g(\bub_{\rm iv},\bvc_{\rm iv})}{h(\bvc_{\rm i},\bub_{\rm i})},
 \ee
where we introduced new functions $\hat r_1(\ub_j)$ and $\hat r_3(\vc_j)$ through the following
equations:
\be{hr1}
r_1(\ub_j)=\hat r_1(\ub_j)\;\frac{f(\ub_j,\bub_j)}{f(\bub_j,\ub_j)}f(\bvb,\ub_j),\qquad
r_3(\vc_j)=\hat r_3(\vc_j)f(\vc_j,\buc).
\ee
For the two other functions we have
\begin{multline}\label{Lu}
\mathcal{L}_a^{(u)}(\{\bub_{\st},\bvc_{\so}\}|\buc)=\sum_{\bub_{\st}\Rightarrow\{\bub_{\rm ii},\bub_{\rm iii}\}}
K_a(\{\bub_{\rm ii}-c,\bvc_{\so},\bub_{\rm iii}\}|\buc)\\
\times \varkappa^{a-k_{\rm ii}}(-1)^{k_{\rm ii}}\frac{r_1(\bub_{\rm ii})}{f(\bvc,\bub_{\rm ii})}f(\buc,\bub_{\rm ii})f(\bub_{\rm iii},\bub_{\rm ii})
f(\bvc_{\so},\bub_{\rm ii}),
\end{multline}
and
\be{Lv}
\mathcal{L}_b^{(v)}(\bvc_{\st}|\bub)=\sum_{\bvc_{\st}\Rightarrow\{\bvc_{\rm ii},\bvc_{\rm iii}\}}
(-1)^{n_{\rm ii}}\frac{r_3(\bvc_{\rm ii})}{f(\bvc_{\rm ii},\bub)}.
\ee
We would like to stress that passing from \eqref{Sab-4} to \eqref{Sab-fact} we did not make any transforms. One can check that substituting
\eqref{G}, \eqref{Lu}, and \eqref{Lv} into \eqref{Sab-fact} we turn back to \eqref{Sab-4}.

The sums over partitions in \eqref{G}, \eqref{Lu}, and \eqref{Lv} can be easily calculated. The most simple is the sum \eqref{Lv}
\be{Lv-res}
\mathcal{L}_b^{(v)}(\bvc_{\st}|\bub)=\prod_{\vc_j\in\bvc_{\st}}
\left(1-\frac{r_3(\vc_{j})}{f(\vc_{j},\bub)}\right).
\ee
If we introduce
 \be{phi-def}
 \varphi(z)=1-\frac{r_3(z)}{f(z,\bub)},
 \ee
 and extend our convention on the shorthand notation to this function, then
\be{Lv-res-1}
\mathcal{L}_b^{(v)}(\bvc_{\st}|\bub)=\varphi(\bvc_{\st}).
\ee

The sum \eqref{G} also is quite simple, because actually this is the Laplace expansion of the determinant of the sum of two matrices
(see appendix~\ref{A-LF} for more details).
Indeed, it is enough to present (see \eqref{gh-C})
\be{gh}
\begin{aligned}
&g(\bub_{\rm iv},\bvc_{\rm iv})=\Delta_{n_{\rm iv}}(\bvc_{\rm iv})\Delta'_{n_{\rm iv}}(\bub_{\rm iv})
\det_{n_{\rm iv}}\Bigl(g(\bub_{{\rm iv}_k},\bvc_{{\rm iv}_j})\bigr),\\
&\frac{1}{h(\bvc_{\rm i},\bub_{\rm i})}= \Delta_{n_{\rm i}}(\bvc_{\rm i})\Delta'_{n_{\rm i}}(\bub_{\rm i})
\det_{n_{\rm i}}\Bigl( \frac{1}{h(\bvc_{{\rm i}_j},\bub_{{\rm i}_k})}\Bigr),
\end{aligned}
\ee
and we immediately recognize the Laplace formula in \eqref{G}. Thus,
\be{G-res}
G_{n_{\so}}(\bub_{\so}|\bvc_{\so})=\Delta_{n_{\so}}(\bvc_{\so})\Delta'_{n_{\so}}(\bub_{\so})\det_{n_{\so}}
\Bigl(g(\ub_{{\so}_k},\vc_{{\so}_j}) +H(\ub_{{\so}_k},\vc_{{\so}_j})\Bigr),
\ee
where
\be{H}
H(\ub_{k},\vc_{j})=\frac{\hat r_3(\vc_{j})\hat r_1(\ub_{k})}{\varkappa h(\vc_{j},\ub_{k})}.
\ee

Finally, the sum \eqref{Lu} can be computed via Lemma~\ref{Long-Det}. Namely,
if we set in \eqref{SumDet1}: $m=a$, $\bar \xi=\buc$, $\bar w=\{\bub_{\st},\;\bvc_{\so}\}$ and
 \be{C1C2}
 C_{1}(w)=\frac{-r_1(w)}{f(\bvc,w)}, \qquad C_{2}(w)=\varkappa,
 \ee
then we obtain equation \eqref{Lu}. Indeed, in this case one has $C_{1}(\vc_j)=0$ due to the product
$1/f(\bvc,w)$ in \eqref{C1C2}. Hence, we automatically have $\bvc_{\so}\subset \bar w_{\st}$, otherwise
the corresponding contribution to the sum vanishes. This means that
when splitting the set $\bar w=\{\bub_{\st},\;\bvc_{\so}\}$
into two subsets we actually should consider only the partitions of the set $\bub_{\st}$ into $\bub_{\rm ii}$ and
$\bub_{\rm iii}$, as we have in \eqref{Lu}.   We
obtain
 \be{ALa}
 \mathcal{L}_a^{(u)}(\bar w|\buc) =\Delta'_a(\buc)\Delta_a(\bar w)
 \det_a\Bigl(\mathcal{M}(\uc_j,w_k)\Bigr),
 \ee
with
 \be{ANa}
 \mathcal{M}(\uc_j,w_k)=(-1)^{a-1}\frac{r_1(w_k)}{f(\bvc,w_k)}t(\uc_j,w_k)h(\buc,w_k) +\varkappa
  t(w_k,\uc_j)h(w_k,\buc),
 \ee
and $\bar w=\{\bub_{\st},\;\bvc_{\so}\}$.

\subsection{Final summation over the partitions of $\bub$ and $\bvc$}

Let $\bx=\{\bub,\bvc\}$. Consider a partition of $\bx$ into subsets $\bx_{\so}$ and $\bx_{\st}$. Let
$\bx_{\so}=\{\bub_{\st},\bvc_{\so}\}$ and $\bx_{\st}=\{\bub_{\so},\bvc_{\st}\}$. Then \eqref{Sab-fact} can be written
in a relatively compact form
 \begin{multline}\label{Sab-6}
S_{a,b}=(-1)^{b}\sum_{\bx\Rightarrow\{\bx_{\so},\bx_{\st}\}}
    h(\bx_{\st},\bub)  g(\bvb,\bx_{\st})\varphi(\bx_{\st})  g(\bx_{\st},\bx_{\so})
    \\
  \times
 \frac{\Delta_{n_{\so}}(\bvc_{\so})\Delta'_{n_{\so}}(\bub_{\so})}{g(\bub_{\so},\bvc_{\so})}
 \det_{n_{\so}} \Bigl(g(\ub_{{\so}_k},\vc_{{\so}_j}) +H(\ub_{{\so}_k},\vc_{{\so}_j})\Bigr)
 \;\Delta'_a(\buc)\Delta_a(\bx_{\so})
 \det_a\Bigl(\mathcal{M}(\uc_j,x_{{\so}_k})\Bigr).
 \end{multline}
Here we have used $\varphi(\ub_j)=1$.

Our goal is to reduce \eqref{Sab-6} to an equation of the following type:
\be{AB}
\sum_{\bx\Rightarrow\{\bx_{\so},\bx_{\st}\}} g(\bx_{\st},\bx_{\so})
\Delta_a(\bx_{\so})
 \det_a\bigl(A_j(x_{{\so}_k})\bigr)\; \Delta_b(\bx_{\st})
 \det_b\bigl(B_j(x_{{\st}_k})\bigr)=\Delta_{a+b}(\bx)\det_{a+b}\begin{pmatrix}
A_j(x_k)\\-~-~-\\B_j(x_k)\end{pmatrix}.
\ee
Here in the r.h.s. we have a matrix consisting of two parts: the entries in the first $a$ rows are $A_j(x_k)$, while in the
remaining rows one has $B_j(x_k)$.

Looking at \eqref{Sab-6} we see that we can set $A_j(x_k)=\mathcal{M}(\uc_j,x_{k})$. We also have a product
$g(\bx_{\st},\bx_{\so})$. The products $ h(\bx_{\st},\bub)$,  $g(\bvb,\bx_{\st})$, and $\varphi(\bx_{\st})$ can be easily absorbed into the determinant
of the matrix $B_j(x_k)$. It remains to construct this matrix $B_j(x_k)$.

Consider a function $F_b(\bz|\bvc)$ depending on $b$ variables $\bz$ and  $b$ variables $\bvc$
\be{F}
F_b(\bz|\bvc)=\Delta'_b(\bz)\Delta_b(\bvc)\det_b\bigl(B_j(z_k)\bigr),
\ee
where
\be{B}
B_j(z_k)=\frac{g(z_{k},\vc_{j}) +H(z_k,\vc_{j})}
{g(z_{k},\bvc)}.
\ee
Obviously, $F_b(\bz|\bvc)$ is a symmetric function of $\bz$ and  a symmetric function of  $\bvc$. Let
$\bz=\bx_{\st}=\{\bub_{\so},\bvc_{\st}\}$. Due to the symmetry of $F_b(\bz|\bvc)$ we can say that the
parameters $\bub_{\so}$ correspond to the first $n_{\so}$ columns of the matrix $B_j(z_k)$,
i.e. $z_k=\ub_{{\so}_k}$ for $k=1,\dots,n_{\so}$. Then in
the remaining columns we should set $z_{n_{\so}+k}=\vc_{{\st}_k}$, $k=1,\dots,b-n_{\so}$. It is easy to see that
in these last columns
\be{B1}
B_j(\vc_{{\st}_k})=\delta_{jk}\prod_{\substack{\ell=1\\ \vc_\ell\ne \vc_{{\st}_k}}  }^b \frac{1}{g(\vc_{{\st}_k},\vc_\ell)},\qquad
k=1,\dots,b-n_{\so},
\ee
where $\delta_{jk}=1$ if $\vc_j=\vc_{{\st}_k}$, and $\delta_{jk}=0$ otherwise.
Thus, the determinant reduces to the determinant of the matrix of the size $n_{\so}\times n_{\so}$. Simple calculation shows that
\be{F-G}
F_b(\{\bub_{\so},\bvc_{\st}\}|\bvc)=\frac{\Delta_{n_{\so}}(\bvc_{\so})\Delta'_{n_{\so}}(\bub_{\so})}{g(\bub_{\so},\bvc_{\so})}
 \det_{n_{\so}} \Bigl(g(\ub_{{\so}_k},\vc_{{\so}_j}) +H(\ub_{{\so}_k},\vc_{{\so}_j})\Bigr),
 \ee
which is exactly the expression in \eqref{Sab-6}. Thus, we recast \eqref{Sab-6} as follows:
 \begin{multline}\label{Sab-7}
S_{a,b}=(-1)^{b}\Delta'_b(\bvc)\Delta'_a(\buc)\sum_{\bx\Rightarrow\{\bx_{\so},\bx_{\st}\}}
      g(\bx_{\st},\bx_{\so})\Delta_b(\bx_{\st})\Delta_a(\bx_{\so})
    \\
  \times  \det_{b} \Bigl(h(x_{{\st}_k},\bub)  g(\bvb,x_{{\st}_k})\varphi(x_{{\st}_k})B_j(x_{{\st}_k})\Bigr)
 \; \det_a\Bigl(\mathcal{M}(\uc_j,x_{{\so}_k})\Bigr).
 \end{multline}
It remains to use \eqref{AB} and we end up with
\be{Sab-8}
S_{a,b}=\Delta'_b(\bvc)\Delta'_a(\buc)\Delta_{a+b}(\bx)\det_{a+b}\begin{pmatrix}
\mathcal{M}(\uc_j,x_{k})\\-~-~-~-~-~-~-~-~-~-~-~-\\-h(x_{k},\bub)  g(\bvb,x_{k})\varphi(x_{k})B_j(x_k)\end{pmatrix},
\ee
where $\bx=\{\bub,\bvc\}$. Substituting here $\mathcal{M}(\uc_j,x_{k})$  and $B_j(x_k)$ we arrive at the statement of Proposition~\ref{prop:main}.

\section*{Conclusion}

In this paper we obtained a determinant representation for the scalar product of semi-on-shell Bethe vectors in
models with $\mathfrak{gl}(2|1)$ symmetry.  When specializing to on-shell (twisted) Bethe vectors, this representation provides determinant formulas
for the form factors of diagonal entries $T_{ii}(u)$ of the monodromy matrix.
 From this result, using the zero modes method \cite{PakRS15a},
one can obtain similar formulas for the form factors of all operators $T_{ij}(u)$. This will be the subject of our forthcoming publication.

It is interesting to compare determinant representations for the scalar products in the models described by different algebras.
In the models with $\mathfrak{gl}(2)$ symmetry a determinant formula exists, if one of the vectors is on-shell \cite{Sla89}. Hereby, the second one remains a generic Bethe vector. The case of $\mathfrak{gl}(3)$-invariant $R$-matrix is more restrictive. For today, only a determinant representation for the
scalar product between on-shell vector and twisted on-shell vector is known \cite{BelPRS12b}. At the same time, in the models with
$\mathfrak{gl}(1|1)$ symmetry a determinant formula for the scalar product exists in the case of generic Bethe vectors \cite{HutLPRS16c},
while in the $\mathfrak{gl}(2|1)$ case it is enough to deal with semi-on-shell Bethe vectors. Thus, the case of superalgebras looks less restrictive.
Therefore, it is possible that determinant formulas for the scalar products of Bethe vectors can be found in the superalgebras of higher rank.

Our results can be applied to the models  with $\mathfrak{gl}(1|2)$ as well. This can be done due to an isomorphism between the
Yangians of $\mathfrak{gl}(2|1)$ and $\mathfrak{gl}(1|2)$ superalgebras \cite{PakRS16a}. In particular, in order to obtain a determinant
representation for the scalar product of $\mathfrak{gl}(1|2)$ semi-on-shell Bethe vectors it is enough to make the replacements
$\bucb\leftrightarrow\bvcb$, $a\leftrightarrow b$, and $r_1\leftrightarrow r_3$ in the formulas for the scalar product
of $\mathfrak{gl}(2|1)$ semi-on-shell Bethe vectors.

Knowing compact determinant representations for form factors of the monodromy matrix entries we can immediately find form factors
of local operators in models for which the solution of the quantum inverse scattering problem is known \cite{KitMT99,MaiT00,GomK00}.
In particular, our results have direct relation to the well known t-J model \cite{ZhaR88} which plays an important role in the condensed matter
physics \cite{ZhaR88,And90}. For special values of the coupling constants the t-J model was studied by the Bethe ansatz in various papers (see e.g. \cite{Sch87,For89,EssK92,FoeK93,Gom02} and references therein). The determinant formulas obtained in the present paper can be directly used for calculating form factors and correlation functions in this model.

It was shown in \cite{PakRS15c} that form factors local operators of $\mathfrak{gl}(3)$-based models can be related to the form factors of the monodromy matrix entries even without the use of the quantum inverse scattering problem. Most probably this relationship exists in the $\mathfrak{gl}(2|1)$ case as well.
We are planning to study this question in our further publications.

\section*{Acknowledgements}
N.A.S. thanks LAPTH in Annecy-le-Vieux for the hospitality and stimulating scientific atmosphere, and CNRS for partial financial support.
The work of A.L.  has been funded by the  Russian Academic Excellence Project 5-100 and by joint NASU-CNRS project F14-2016.
The work of S.P. was supported in part by the RFBR grant 16-01-00562-a.
N.A.S. was  supported by  the grants RFBR-15-31-20484-mol-a-ved and RFBR-14-01-00860-a.

\appendix

\section{Summation of rational functions\label{A-CI}}

\subsection{Single sums\label{A-CI1}}

Here we give an example of the derivation of identities \eqref{sum-rows}.

Consider a contour integral
\be{con-int}
I=\frac1{2\pi i}\oint_{|z|=R\to\infty}\frac{dz}{x_k-z}\prod_{l=1}^b\frac{z-\vb_l}{z-\vc_l}\,.
\ee
Taking the residue at infinity we find that $I=-1$. On the other hand, this integral is equal to the sum of the residues within the
integration contour. Hence,
\be{sumres}
-1=-\prod_{l=1}^b\frac{x_k-\vb_l}{x_k-\vc_l}+\sum_{j=1}^b \frac{1}{x_k-\vc_j}\,\frac{\prod_{l=1}^b (\vc_j-\vb_l)}
{\prod_{l=1,\; l\ne j}^b (\vc_j-\vc_l)}\,.
\ee
Rewriting everything in terms of the function $g$ we obtain
\be{sum-rows3}
\sum_{j=1}^b g(x_k,\vc_j)\Omega_{a+j}=\frac{g(x_k,\bvc)}{g(x_k,\bvb)}-1.
\ee
This is one of the identities in \eqref{sum-rows}. All the other identities can be proved exactly in the same way.

\subsection{Multiple sums\label{A-CI2}}

\subsubsection{Laplace formula\label{A-LF}}

Let $\#\bu=\#\bv=n$. Let $A$ and $B$ be $n\times n$ matrices whose matrix elements are indexed by the parameters $u_j$ and $v_k$:
$A_{jk}=A(u_j,v_k)$ and $B_{jk}=B(u_j,v_k)$.  The Laplace formula  gives an expression of $\det(A+B)$ in terms
of $\det A$ and $\det B$:
\be{Laplac}
\det\bigl(A(u_j,v_k)+B(u_j,v_k)\bigr)=\sum (-1)^{[P_u]+[P_v]}\det\bigl(A(u_{{\so}_j},v_{{\so}_k})\bigr)
\det\bigl(B(u_{{\st}_j},v_{{\st}_k})\bigr).
\ee
The sum is taken over partitions $\bu\Rightarrow\{\bu_{\so},\bu_{\st}\}$ and $\bv\Rightarrow\{\bv_{\so},\bv_{\st}\}$  with the restriction $\#\bu_{\so}=\#\bv_{\so}$. Recall that according to our convention the elements of every subset are ordered in the natural order.
 $[P_u]$ denotes the parity of the permutation mapping the union $\{\bu_{\so},\bu_{\st}\}$ into the naturally ordered set $\bu$. The notation $[P_v]$ has an analogous meaning. Equivalently, one can say that $[P_u]+[P_v]$ is the parity of the permutation mapping the sequence of the subscripts of
the union $\{\bu_{\so},\bu_{\st}\}$ into  the sequence of the subscripts of
the union $\{\bv_{\so},\bv_{\st}\}$.

Let us introduce two functions
\be{AB0}
\mathcal{A}(\bu|\bv)=\Delta_n(\bu)\Delta'_n(\bv)\det A(u_j,v_k)\mb{and}
\mathcal{B}(\bu|\bv)=\Delta_n(\bu)\Delta'_n(\bv)\det B(u_j,v_k).
\ee
These functions  depend on two sets of variables $\bu$ and $\bv$. They are symmetric over $\bu$ and symmetric over $\bv$.
Then equation \eqref{Laplac} can be written in the following form:
\be{Laplac1}
\Delta_n(\bu)\Delta'_n(\bv)\det\bigl(A(u_j,v_k)+B(u_j,v_k)\bigr)=\sum \mathcal{A}(\bu_{\so}|\bv_{\so})\mathcal{B}(\bu_{\st}|\bv_{\st})
g(\bu_{\st},\bu_{\so})g(\bv_{\so},\bv_{\st}).
\ee
Indeed, multiplying \eqref{Laplac} with $\Delta_n(\bu)\Delta'_n(\bv)$ and using obvious relations
\be{Del-Del}
\begin{aligned}
\Delta_n(\bu)&= (-1)^{[P_u]}\Delta_{n_{\so}}(\bu_{\so})\Delta_{n-n_{\so}}(\bu_{\st})g(\bu_{\st},\bu_{\so}),\\
\Delta'_n(\bv)&= (-1)^{[P_v]}\Delta'_{n_{\so}}(\bv_{\so})\Delta'_{n-n_{\so}}(\bv_{\st})g(\bv_{\so},\bv_{\st}),
\end{aligned}
\ee
we arrive at \eqref{Laplac1}.

In section~\ref{S-CSP} we use \eqref{Laplac1} in the particular case of Cauchy determinants. For completeness, we recall
that for arbitrary complex $\bu$ and $\bv$ with $\#\bu=\#\bv=n$ the Cauchy determinant is defined as
\be{Cauchy1}
C_n=\det\left(\frac1{u_j-v_k}\right).
\ee
It has an explicit presentation in terms of double products
\be{Cauchy2}
C_n=\frac{\prod_{1\le k<j\le n}(u_j-u_k)(v_k-v_j)}{\prod_{j=1}^n\prod_{k=1}^n (u_j-v_k)}.
\ee
From this we immediately obtain
\be{gh-C}
g(\bu,\bv)=\Delta_n(\bu)\Delta'_n(\bv)\det\bigl(g(u_j,v_k)\bigr), \qquad
\frac1{h(\bu,\bv)}=\Delta_n(\bu)\Delta'_n(\bv)\det\left(\frac1{h(u_j,v_k)}\right).
\ee

\subsubsection{Other sums over partitions\label{A-OSP}}
 In the core of the proof, we use different equalities, that were proven elsewhere. We recall them in the present appendix.
\begin{lemma}\label{main-ident-C}
Let $\bw$, $\bu$ and $\bv$ be sets of complex variables with $\#\bu=m_1$,
$\#\bv=m_2$, and $\#\bw=m_1+m_2$. Then
\begin{equation}\label{Sym-Part-new1}
  \sum
 g(\bw_{\so},\bu)g(\bw_{\st},\bv)g(\bw_{\st},\bw_{\so})
 = \frac{g(\bw,\bu)g(\bw,\bv)}{g(\bu,\bv)}\,,
 \end{equation}
 where the sum is taken with respect to all partitions of the set $\bw$ into
subsets $\bw_{\so}$ and $\bw_{\st}$ with $\#\bw_{\so}=m_1$ and $\#\bw_{\st}=m_2$.
\end{lemma}
The proof of this Lemma is given in \cite{Sla16}.

\begin{lemma}\label{main-ident}
Let $\bw$, $\bu$ and $\bv$ be sets of complex variables with $\#\bu=m_1$,
$\#\bv=m_2$, and $\#\bw=m_1+m_2$. Then
\begin{equation}\label{Sym-Part-old1}
  \sum
 K_{m_1}(\bw_{\so}|\bu)K_{m_2}(\bv|\bw_{\st})f(\bw_{\st},\bw_{\so})
 = (-1)^{m_1}f(\bw,\bu) K_{m_1+m_2}(\{\bu-c,\bv\}|\bw).
 \end{equation}
The sum is taken with respect to all partitions of the set $\bw$ into
subsets $\bw_{\so}$ and $\bw_{\st}$ with $\#\bw_{\so}=m_1$ and $\#\bw_{\st}=m_2$.
\end{lemma}
The proof of this Lemma is given in \cite{BelPRS12a}.

\begin{lemma}\label{Long-Det}
Let $\bar w$ and $\bar\xi$ be two sets of generic complex numbers with $\#\bar w=\#\bar\xi=m$. Let
also $C_{1}(w)$ and $C_{2}(w)$ be two arbitrary functions of a complex variable $w$. Let us extend our convention on the
shorthand notation to the products of these functions. Then
\begin{multline}\label{SumDet1}
\sum K_m(\bar w_{\so}-c, \bar w_{\st}|\bar \xi)f(\bar \xi, \bar w_{\so})f(\bar w_{\st},\bar w_{\so})
C_{1}(\bar w_{\so})C_{2}(\bar w_{\st})\num
=\Delta'_m(\bar\xi)\Delta_m(\bar w)
\det_m\Bigl(C_{2}(w_k)t(w_k,\xi_j)h(w_k,\bar\xi)+(-1)^m C_{1}(w_k)t(\xi_j,w_k)h(\bar\xi,w_k)\Bigr).
\end{multline}
Here the sum is taken over all possible partitions of the set $\bar w$ into subsets $\bar w_{\so}$
and $\bar w_{\st}$.
\end{lemma}
The proof of this Lemma is given in \cite{BelPRS12b}.


\end{document}